\documentclass[12pt]{article}
\usepackage{graphicx}
\newcommand{\beq}{\begin{equation}}
\newcommand{\eeq}{\end{equation}}
\newcommand{\bea}{\begin{eqnarray}}
\newcommand{\eea}{\end{eqnarray}}

\def\fun#1#2{\lower3.6pt\vbox{\baselineskip0pt\lineskip.9pt
  \ialign{$\mathsurround=0pt#1\hfil##\hfil$\crcr#2\crcr\sim\crcr}}}
\begin{document}
\begin{titlepage}
\begin{flushleft}
       \hfill                       FIT HE - 04-01 \\
\end{flushleft}
\vspace*{3mm}
\begin{center}
{\bf\LARGE Stability of RS brane for tachyonic scalars\\ }
\vspace*{5mm}

{\large Kazuo Ghoroku\footnote{\tt gouroku@dontaku.fit.ac.jp}\\ }
\vspace*{2mm}
{
${}^2$Fukuoka Institute of Technology, Wajiro, Higashi-ku}\\
{
Fukuoka 811-0295, Japan\\}
\vspace*{5mm}

\vspace*{10mm}

\end{center}

\begin{abstract}
We study the stability of the RS brane embedded in the AdS$_5$ vacuum of 5d
gauged supergravity, where many tachyonic scalars exist. We consider a model
in which these scalars couple to the brane such that the BPS conditions 
are satisfied to preserve the bulk supersymmetric configuration. 
In this case, we find that these tachyons are not
trapped on the brane and only the massless dilaton is localized. As a result,
the braneworld is stable. Further, the effective action of trapped fields 
is studied by using the brane running method developed recently, and we 
find that the action is independent of the brane position. 

\end{abstract}
\end{titlepage}

\section{Introduction}

It is an interesting idea to consider our 4d world as
a thin three-brane embedded in $AdS_5$ space-time as proposed by 
Randall-Sundrum (RS)~\cite{RS1,RS2}. 
This bulk space is considered as an extended
part of $AdS_5\times S^5$, which is realized near the accumulated D3 branes
in the type IIB superstring theory. 
In this context, the bulk theory can be supposed as the 
maximally symmetrized 5d gauged supergravity with
massive Kaluza-Klein (KK) modes of $S^5$.
In this theory, many tachyonic scalars are included, and they play an
important role \cite{GPPZ,PW,BS} in the context of AdS/CFT correspondence 
\cite{M1,GKP1,W1,Poly1}. Since their masses
are within the Bleitenrhoner-Freedman bound \cite{BF}, then the bulk is stable.
While, it is known that the bulk tachyons can be trapped on the brane as 4d
tachyons when the tachyons are free and the brane action is given
by the tension only \cite{GN}. 
Then the braneworld is unstable in such a case. 

\vspace{.2cm}
In our present model, the brane is embedded in the AdS$_5$ in such a way
to keep the BPS conditions \cite{ST,DFGK} 
and to preserve the bulk supersymmetric
configurations. 
This is realized by a special form of the brane
action or the brane-scalar coupling \cite{BKP,G1},
then the situation for the localization is non-trivially changed. 
So, it is important to make clear the localization problem
in the present model based on the gauged supergravity.
Our purpose is to examine the stability of the braneworld
through the investigation of the localization for tachyons
in our brane model. It would be necessary to make clear
this point when we try to construct the braneworld based on 
the superstring theory compactified to AdS$_5\times S^5$.

\vspace{.2cm}
Another point to be studied here is the effective action of
the brane. In the original RS model,
the brane action is simply expressed by a tension parameter
only. In the present case, however,
it is given as a function of scalar fields 
as mentioned above. 
So we might need some principle what kind of brane action we choose.
The most important factor would be the stability of the braneworld. 
Since the form of the brane action is deeply related to the
localization of the bulk fields, then it also controls the stability of
the braneworld.

However, in general, the action is not invariant for the shift of the
brane position even if a form of the action is fixed at some point.
Actually we need
various kinds of terms in the brane action when the brane position is
changed by keeping the background configuration
and its fluctuation modes to be unchanged \cite{LMS,LR,Re,U,BGY}.
This procedure is called as brane running and it would provide a
renormalization group flow of the brane action in
the following sense.
The fields on the brane could
couple to the bulk modes, and interact with them. From the viewpoint of
AdS/CFT, this interaction can be interpreted as the one with the cutoff 
CFT living on the brane. Through this interaction, the brane action would 
receive cutoff dependent corrections. As a result, 
the parameters of the brane action are running and they varies 
according to the
flow equations obtained by the brane running method.
In other words, the parameters of the brane action on a different
position are related each other by these flow equations.
Then a simple brane action of RS model is regarded
as the one obtained under a special renormalization condition at a 
an appropriate
renormalization point. So it might be changed to a more complicated or 
an unstable form of action when
the renormalization point is shifted even if the original form is simple
and stable.

According to this idea, we examine the running behavior of the brane action 
for a scalar field, which is trapped on the brane, and
derive the effective action. Through this
action, we can see the effect on the brane from the bulk of
gauged supergravity and also get an aspect for the stability of the system.

\vspace{.2cm}
In Section 2, the model used here is set, and the localization of
scalar fields and the stability of the braneworld are examined in Section 3. 
In Section 4, the effective action for the brane is obtained by the brane
running method, and the effect of the bulk on the brane is discussed.
Concluding remarks are given in the final section.

\section{Setting of the model}

As a bulk action,
consider the bosonic part of a truncated 5d gauged supergravity 
\beq
   S_{\rm g}=\int d^4\!xdy\sqrt{-g}
   \left\{{1\over 2\kappa^2}R 
    -{1\over 2}\sum_I(\partial\phi_I)^2-V(\phi)\right\}
       +{1\over 2\kappa^2}2\int d^4x\sqrt{-g}K \ ,
                                                     \label{ac1g}
\eeq
where $K$ is  the extrinsic curvature on the boundary, and
the potential $V$ of scalars $\phi_I$
is written by their superpotential $W(\phi_I)$ as
\beq
 V={v^2\over 8}\sum_I\left({\partial W\over \partial \phi_I}\right)^2
   -{v^2\over 3}W^2 . \label{superPot}
\eeq
The gauge coupling parameter $v$ is fixed 
from the AdS$_5$ vacuum \cite{FGPW} 
by fixing the radius of AdS as a unit length.
And we are considering in the
Einstein frame \footnote{
Here we take the following definition, $R_{\nu\lambda\sigma}^{\mu}
=\partial_{\lambda}\Gamma_{\nu\sigma}^{\mu}-\cdots$, 
$R_{\nu\sigma}=R_{\nu\mu\sigma}^{\mu}$ and $\eta_{AB}=$diag$(-1,1,1,1,1)$. 
Five dimensional suffices are denoted by capital Latin and four
dimensional
ones by  Greek letters.
}.
The other ingredient is the brane action, which is given here as \cite{G1}
\beq
    S_{\rm b} = -{v}\int d^4x dy\sqrt{-g}W(\phi_I)\delta(y-y_h). \label{baction}
\eeq
The reason for choosing of this form is made clear in the followings, and the
brane position $y_h$ is set as $y_h=0$ hereafter for simplicity.

\vspace{.3cm}
The background solutions are obtained under the ansatz, $\phi_I=\phi_I(y)$
and 
\beq
 ds^2= A^2(y)\eta_{\mu\nu}dx^{\mu}dx^{\nu}
           +dy^2  \, . \label{metrica}
\eeq
Here $\eta_{\mu\nu}=$ diag(-1,1,1,1) and
the coordinates parallel to the brane are denoted by $x^{\mu}=(t,x^i)$,
$y$ being the coordinate transverse to the brane. 
In this set up, the four dimensional slice perpendicular to $y-$axis is 
the Minkowsky space-time, and
the BPS solutions for the bulk configuration
are obtained by solving the following first order equations \cite{ST},
\beq
 \phi_I'={v\over 2}{\partial W\over\partial\phi_I}, \qquad
    {A'\over A}=-{v\over 3}W,  \label{first-order}
\eeq
where $'=d/dy$. The solutions of (\ref{first-order}) satisfy the equations 
of motion of $S_{\rm g}$ as is well-known, and they preserve supersymmetry
\cite{ST,DFGK} in the bulk. Further, the equations 
(\ref{first-order}) at the brane position coincide with
the boundary conditions in solving the equations of motion
of the total action, $S=S_{\rm g}+S_{\rm b}$, iff $S_{\rm b}$ was
given by the form of (\ref{baction}).
This is the reason why the brane action is given as (\ref{baction}).
Then, in this case, the brane can be embedded 
at any point of $y$
since the boundary conditions are satisfied at any $y$ for the solutions of
(\ref{first-order}). Another point to be noticed is that this system could be
extended to a supersymmetric braneworld solutions by adding other
necessary terms on the brane \cite{BKP}.

\vspace{.3cm}
As for the scalar fields in the
5d gauged supergravity, there exist many tachyonic fields and they play
an important role in the gauge/gravity correspondence since they
couple to the relevant operators of $\mathcal{N}=4$ SYM as deformations 
of the CFT on the boundary. 
When we consider a free tachyonic scalar which
does not couple to the brane, this scalar is trapped on the brane
as a 4d tachyonic scalar \cite{GN}. In this case, the braneworld 
destabilizes and
the existence of such a tachyon in the bulk would be rejected.
In the present case, however, the situation is different since
the the scalars couple to the brane through $W(\phi)$ in a special form. 
We can see in the next section that this coupling changes the situation 
of the stability for braneworld. 


\section{Scalar locarization}

In order to make clear the issue, we consider here a simple
AdS$_5$ solution, $A=e^{-|y|}$ and $\phi_I=0$. Here we 
set the parameters as $\kappa^2=2$ and $v=-2$ for simplicity. 
According to \cite{GN}, we examine the localization of scalars
in terms of the linearized equations. 
For simplicity, consider the case of one scalar, $\phi$, which is 
expanded as $\phi=\bar{\phi}+\chi$ around its classical solution
$\bar{\phi}(=0)$. 
So the equation of $\chi$ is obtained as 
\beq
  {\chi}''+4{A'\over A}{\chi}'
           +{m^2\over A^2}\chi=
\left({\partial^2 V(0)\over \partial \phi^2}
       -2{\partial^2 W(0)\over \partial \phi^2}\delta(y)\right)\chi,
                         \label{chi-eq1}
\eeq
where $m^2$ is the four dimensional mass square of $\chi$.
Due to the second term of the right hand side in (\ref{chi-eq1}),
the following boundary condition is needed,
\beq
  \chi'(0)=-{\partial^2 W(0)\over \partial \phi^2}\chi(0). \label{boundaryc}
\eeq
Hereafter we denote as $V^{(i)}={\partial^i V(0)\over \partial \phi^i}$
and $W^{(i)}={\partial^i W(0)\over \partial \phi^i}$, 
then the equation (\ref{chi-eq1}) can be rewritten into the following form,
\beq
 [-\partial_z^2+V(z)]u(z)=m^2 u(z) , \ \label{warp3}
\eeq
\beq
 V(z)={9\over 4}(\partial_yA)^2+{3\over 2}A\partial_y^2A+A^2V^{(2)}
-2 W^{(2)}\delta(y),
\eeq 
where we introduced $u(z)$ and $z$ defined as 
$u=A^{-3/2}\chi$ and $\partial z/\partial y=\pm A^{-1}$.

For $A=e^{- |y|}$, the bound state wave function is solved as
$u=x^{1/2}K_{\nu}(|m|x)$ where $x=|z|+1$ and $\nu=\sqrt{4+V^{(2)}}$.
Since the bound state should be restricted to the region $m^2\leq 0$, 
the solution
is written in terms of the absolute value of negative $m^2=-|m|^2$,
and $K_{\nu}(x)$ is the modified Bessel function.
Then the boundary condition (\ref{boundaryc}) is obtained as
\beq
  \left(2+\nu+W^{(2)}\right)K_{\nu}(|m|)=|m|K_{\nu+1}(|m|),
    \label{boundaryc2}
\eeq
where the AdS$_5$ radius is taken as unit and $W^{(0)}=-3/2$ since 
we set as $\kappa^2=2$ and $v=-2$. Further we demanded
$W^{(1)}=0$, which is needed for $\bar{\phi}=0$ and satisfied for all the
known superpotentials. And from (\ref{superPot}), 
\beq
  M^2\equiv V^{(2)}=4W^{(2)}+(W^{(2)})^2, \qquad \nu=|2+W^{(2)}|. 
\eeq
So the boundary condition (\ref{boundaryc2}) depends only on 
the parameter $W^{(2)}$.
For $W^{(2)}>-2$ ($M^2 > -4$), (\ref{boundaryc2}) is written as
\beq
  2\nu={|m|K_{\nu+1}(|m|)\over K_{\nu}(|m|)},  \label{boundaryc3}
\eeq
and this is satisfied only for $m=0$. While for $W^{(2)}\leq -2$, we have
$\nu=-(2+W^{(2)})$, then
$2+\nu+W^{(2)}=0$. In this case, (\ref{boundaryc2}) is satisfied only 
for $\nu=0$ and $m=0$, then 
$W^{(2)}=-2$ and $M^2=-4$. 
After all, we find that there is no tachyonic bound 
state and the trapped state might be 
seen in the case of $M^2\geq -4$, within the BF bound, only for $m=0$.

\vspace{.3cm}
In order to see whether the zero mode, $m=0$, is really localized or not,
we must check over the normalizability of the mode with respect to $y$-
integration. From (\ref{chi-eq1}), the $y$-dependent
part of $\chi$ for $m=0$ is obtained by parametrizing 
as $\chi=\chi_t(y)\chi_b(x)$,
\beq
 \chi_t(y)
          =e^{(2-\nu)y}.
\eeq
From the normalizability of the kinetic term, we demand
\beq
  \int_0^{\infty} dy A^2(y)\chi_t(y)^2=\int_0^{\infty} dy~ e^{2\left(1-\nu
\right)y}<\infty ,
\eeq
then $\nu <1$ or $M^2> -3$.

On the other hand, we must also demand the normalizability of the potential 
$V(\chi)$ simultaneously. In general, however, $V(\chi)$ 
is expressed by infinite power series as 
$V(\chi)=\sum_i^{\infty}V^{(i)}\chi^i/i!$, so we must demand
$2-\nu\leq 0$ for the normalizability of all these terms. 
Then the condition $M^2\geq 0$ is needed. This means that the bulk mass 
square of the trapped state must be non-negative. Then
we can conclude that the bulk tachyonic
scalar ($M^2 < 0$) can not be trapped neither as a tachyon nor as a zero mode.

\vspace{.2cm}
For the dilaton, the potential is zero, $V(\phi)=0$, and
$W(\phi)=0$ \cite{DZ}. The situation for the localization is the same
with the case of a free scalar field which does not couple to the brane.
We find that this field can be trapped as a massless scalar on the brane,
and it does not affect the stability of the brane solution.

\vspace{.3cm}
After all, the RS braneworld is stable against for any tachyonic scalar 
of the 
gauged supergravity when we take the brane action as (\ref{baction}). 
In order to see the necessity of this form for its stability, consider
a small modification of
(\ref{baction}) without changing the background configuration. For example,
considr the following brane action,
\beq
    S_{\rm b} = -{v}\int d^4x dy\sqrt{-g}
  \left(W(\phi)+{1\over 2}\epsilon\phi^2\right)\delta(y-y_h). \label{baction2}
\eeq
with a small parameter $\epsilon$. 
Here the bulk action is not changed. In this case, the bulk AdS$_5$ is 
preserved, but the boundary condition (\ref{boundaryc3}) is changed as
\beq
  2\nu+\epsilon={|m|K_{\nu+1}(|m|)\over K_{\nu}(|m|)}, \label{bundaryc4}
\eeq
and we find that the scalar is trapped as a tachyon for the cases of
$\left\{\nu<1, \epsilon <0\right\}$
and $\left\{\nu\geq 1, \epsilon > 0\right\}$. In general, 
in the gauged supergravity, both kinds of scalars of $\nu<1$ and 
$\nu\geq 1$ are included, then we could not get a stable
brane solutions for a finite value of $\epsilon$. 
Then the stable solution would be
restricted to the case of $\epsilon=0$. This implies that the background
configuration should be changed when the BPS conditions are broken.
In the next section, we study the effective action how these bulk 
scalars are
observed on the brane.

\section{Effective action and brane running}
\vspace{.3cm}
Here we estimate the effective brane action $S^{\rm eff}_b$ to study how
we observe the bulk fields on the brane.
From the viewpoint of 
path-integral formulation, it
can be obtained as \cite{Gidd,GKa,HeSk,GY}
\beq
 S^{\rm eff}_b={1\over 2}S_{\rm b} + \ln Z_5(g,\phi)
  \label{effaction-1}
\eeq
\beq
  Z_5(g,\phi)=\int_{G(x,0),\phi(x,0)|\rm{fixed}} DG D\phi 
   ~{\rm exp}\left(i\int_{y\geq 0}d^4xdyL_{\rm g}\right),   \label{effective-1}
\eeq
where $g$ and $\phi$ represent the boundary value of $G(x,y)$ and 
$\phi(x,y)$ respectively. And
$S_{\rm b}$ and $S_{\rm g}\equiv \int d^4xdyL_{\rm g}$ are defined 
in the previous section. 

As for $\ln Z_5(g,\phi)$, it is related to the 
conformal field theory (CFT) on the
boundary ($y\to -\infty$) in the context of AdS/CFT correspondence. And
it can be estimated by WKB approximation in terms of a
classical solution as
\beq
 \ln Z_5(g)=S_{\rm CT}+S_{\rm CFT}
\eeq
where $S_{\rm CT}$ represent the divergent term at
$y\to -\infty$. The CFT generating functional $S_{\rm CFT}$, which is
finite at $y\to -\infty$,
is obtained by subtracting the divergent $S_{\rm CT}$ from $\ln Z_5(g)$.
However, here, $y$ is taken at some point, $y_0$, off
the boundary $y=-\infty$,
then $S_{\rm CFT}$ represents a "cut-off" CFT and it is written by
the fields at $y=y_0$ not at $y=-\infty$.
In this context, $\ln Z_5(g)$ has been estimated in
terms of the asymptotic expansion of the equation of motion, and
we find the counter term for a scalar of conformal dimension $\Delta$ as
\cite{HSS,BFS}
\beq
 S^{\rm s}_{\rm CT}=\int_{y=y_0} d^4x\sqrt{-g}
      \left({4-\Delta\over 2}\phi^2-{1\over 4(\Delta-3)}g^{\mu\nu}\partial_{\mu}\phi\partial_{\nu}\phi+\cdots\right) , \label{holog}
\eeq
for $\Delta\neq 3$. While
for $\Delta=3$, the coefficient $-{1\over 4(\Delta-3)}$ of the kinetic 
term is replaced by 
$+{1\over 4}\ln(\epsilon)\equiv {y_0/ 2}$, which corresponds to the conformal 
anomaly due to the scalar.
This is seen also in the brane running method as shown below, and this 
kinetic term is canceled
out in the effective action. So the kinetic term of this scalar field
disappears. However the scalar of $\Delta=3$ ($M^2=-3$) is not trapped
as shown in the previous section, so we should consider the above action
is useful only 
for the classical solution as on-shell action 
in the case of $\Delta=3$ scalar.
Here we concentrate on
the scalar whose zero mode is trapped. It is the dilaton of
$\Delta=4$ in the present case. We notice that the same result 
with (\ref{holog}) is obtained for this scalar
simply by integrating the $y$-dependent part in the 5d action as
\beq
 S^{\rm s}_{\rm CT}=\int_{y=y_0} d^4x\sqrt{-g}
      \left(-{1\over 4}g^{\mu\nu}\partial_{\mu}\phi\partial_{\nu}\phi+\cdots\right) ,
\eeq
which is precisely equivalent to (\ref{holog})
for $\Delta=4$.

\vspace{.5cm}
In the next, we estimate the brane action $S_{\rm b}$ at 
an arbitrary point of $y$
by using the brane running method to obtain the effective action for
the trapped scalar field $\phi$.
The scalar is expanded as $\phi=\bar{\phi}+\chi$.
For $\bar{\phi}=0$, the linearized equation for $\chi$ is obtained as 
\beq
  {\chi}''+4{A'\over A}{\chi}'
           +{-q^2\over A^2}\chi=\left({\partial^2 V(0)\over \partial \phi^2}
           -2{\partial^2 W(0)\over \partial \phi^2}\delta(y)\right)\chi,
                         \label{chi-eq}
\eeq
where $q^2$ is the four dimensional momentum square of $\chi$
and we take $y_0=0$ for simplicity. Then the boundary condition for
$\chi$ is written as,
\beq
 {\chi'(0)\over \chi(0)}
=-{\partial^2 W(0)\over \partial \phi^2}|_{y=0}. 
\eeq
We extend this equation to the region, $y>y_0(=0)$, where the running
brane arrived, by introducing the scalar part of running brane action as
\beq
    S_{\rm b}^{\rm (s)} = -\int d^4x \sqrt{-g}\left\{
          \sum_{i=0}{\chi^i\over i!}s_{(i)}(y)
        +{1\over 2}\tau_k(y)(\partial\chi)^2+\cdots\right\}, \label{baction-R}
\eeq
where dots represent other higher derivative and non-local terms. 
Then the extended boundary condition is given as
\beq
  2{\chi'(y)\over \chi(y)}=s_{(2)}(y)-\tau_k(y){-q^2\over A^2(y)}
         +O\left(({q^2\over A^2(y)})^2\right), \label{boundchiy}
\eeq
with the initial values, 
$s_{(2)}(0)=-2{\partial^2 W(0)\over \partial \phi^2}\equiv -2W^{(2)}$ and 
$\tau_k(0)=0$.
By differentiating (\ref{boundchiy}) with respect to $y$ and using 
(\ref{chi-eq}), we obtain the following $\chi$-independent flow equations,
\beq
  {s_{(2)}}'=-{1\over 2}s_{(2)}^2-4{A'\over A}s_{(2)}+2V^{(2)}
             , \label{tau20-eq}
\eeq
\beq
  \tau_k'=-s_{(2)}\tau_k-2{A'\over A}\tau_k+2
             , \label{tauk-eq}
\eeq
where $V^{(2)}\equiv {\partial^2 V(0)\over \partial \phi^2}=(W^{(2)})^2-4W^{(2)}$.
These equations are solved as 
\beq
   s_{(2)}={\rm const.}=-2W^{(2)}, \qquad \tau_k=-{1-e^{2\tilde{\mu}y}\over
\tilde{\mu}},
\eeq 
where $\tilde{\mu}=1+W^{(2)}$. The above solution for
$\tau_k$ is useful for $W^{(2)}\neq -1$. In the case of $W^{(2)}=-1$,
we find $M^2=-3(=V^{(2)})$ and $\Delta=3$. And we obtain
\beq
  \tau_k=2y=\ln(\epsilon) \, , \label{anomaly2}
\eeq
where $\epsilon$ is the cutoff used in \cite{HSS,BFS}.
As mentioned in the previous section, this kinetic term cancels out 
in ${1\over 2}S_{\rm b} + S_{\rm CT}$. However,
as shown in the previous section, the scalar of $M^2<0$ is not trapped, so
we don't consider the effective action for this scalar. However
the above result (\ref{anomaly2}) is very interesting since this reproduces the conformal anomaly which has been shown in the holographic approach 
\cite{HSS,BFS}. So we could see again the anomaly through the brane running method as shown in the
case of pure gravity \cite{BGY}.

\vspace{.3cm}
Using the above results, we can obtain the effective brane action at an
arbitrary
point of running position, or at an arbitrary mass scale of the 4d field
theory. The only trapped scalar here is the dilaton, for 
which $M^2=W^{(2)}=0$, then
\beq
   s_{(2)}=0, \qquad \tau_k=e^{2y}-1.
\eeq 
After all, we arrive at the following effective brane action for the trapped
scalar,
\beq
  S_{\rm eff}^{\rm s}={1\over 2}S_{\rm b}^{\rm (s)}+S^{\rm s}_{\rm CT}
     =\int d^4x\sqrt{-\hat{g}}
      \left(-{1\over 2}\hat{g}^{\mu\nu}\partial_{\mu}
\phi\partial_{\nu}\phi+\cdots\right),
\eeq
where the metric $\hat{g}^{\mu\nu}(x)$ is defined as 
$ds^2=A^2(y)\hat{g}_{\mu\nu}(x)dx^{\mu}dx^{\nu}+dy^2$.
This result implies that the low energy action is independent of the brane
position as shown for the Einstein term in the gravitational case
\cite{BGY}.
As a result, we have
an effective brane action which includes the Einstein term and the dilaton as
\beq
  S^{\rm eff}_b = \int d^4x\sqrt{-\hat{g}}\left(
      {1\over 4\mu\kappa^2}\hat{R}-{1\over 4}\hat{g}^{\mu\nu}\partial_{\mu}
\phi\partial_{\nu}\phi+\cdots \right)
      +S_{\rm CFT},
   \label{EactionRS}
\eeq
Except for the higher derivative terms, 
the action is independent of the energy-scale $y$. In other words,
we can see the same low energy theory at any mass scale.
This would be the reflection of the conformal invariance of the AdS bulk.
And all the tachyonic scalars are living in the bulk and they affect the
fields on brane through $S_{\rm CFT}$, but the above low energy part is
not affected.

\section{Concluding remarks}

The stability of the RS braneworld is examined in the AdS$_5$ vacuum of
gauged supergravity. Although there are several tachyonic scalars in this
model, their masses are within the bound of 
Breitenlohner-Friedman. Then the bulk is stable.
While, a tachyonic scalar in the bulk could be
trapped on the brane as a four dimensional tachyon when there is no 
brane-scalar coupling, then the brane becomes
unstable in this case. In the present case, we set a brane-scalar coupling 
in such a way that the supersymmetric bulk
configuration is preserved. In other words, the BPS conditions are
satisfied in the bulk and also on the brane.
Just on the brane position, the BPS conditions are
equivalent to the boundary conditions of the equations of motion. 
Then the brane, in our model,
can be embedded at any point in the bulk when the solutions
are BPS. 

In this case
the situation for the trapping of tachyons
is non-trivial, and we find that the tachyons are not trapped on the brane 
in our model mentioned above. 
Among the scalars in the model, 
only the massless dilaton is trapped. 
Then our braneworld is stable against many tachyonic fields in the bulk.
This stability is
supported by the bulk supersymmetry and the BPS conditions satisfied up
to the brane position. Actually,
we can see that the braneworld becomes unstable for a small modification,
which breaks the BPS conditions, of
the brane action.

\vspace{.3cm}
For the trapped dilaton field,
the effective action is studied by the brane running method to see the effect
from the bulk to the brane action. The action given at some point of the fifth 
coordinate $y$ may represent the one obtained
at the cutoff of $\ln\epsilon=y$. The cutoff dependence of this action
can be seen by shifting
$y$ according to the brane running method. After
performing this method, we could find that the action of relevant terms are
independent of the brane position $y$. This result is similar to the case of the
Einstein action for the trapped graviton. We can say that this y-independence
is the reflection of the bulk supersymmetry
or conformal symmetry of CFT on the brane. We find further the anomaly 
for the kinetic term of the scalar field of conformal dimension $\Delta=3$.
This coincides with the one found in the holographic approach to obtain the on-shell action. 
Then we could assure that
the brane running method would give a correct flow of the parameters
in the brane action.


\section*{Acknowledgments}
This work has been supported in part by the Grants-in-Aid for
Scientific Research (13135223)
of the Ministry of Education, Science, Sports, and Culture of Japan.

\end{document}